\documentclass[aps,showpacs,onecolumn,nofootinbib,byrevtex]{revtex4}
\usepackage{mathrsfs,amsmath,graphicx,natbib,bm}
\preprint{adahsdkh}
\begin{document}
\title{Single polymer dynamics in elongational flow and the confluent Heun equation}
\author{D. Vincenzi}
\email{Dario.Vincenzi@ds.mpg.de}
\affiliation{Max-Planck-Institut f\"ur Dynamik und 
Selbstorganisation, 37077 G\"ottingen, Germany}
\affiliation{School of Mechanical \& Aerospace Engineering
and LASSP, Cornell University, Ithaca, NY 14853, USA}
\author{E. Bodenschatz}
\affiliation{Max-Planck-Institut f\"ur Dynamik und 
Selbstorganisation, 37077 G\"ottingen, Germany}
\affiliation{School of Mechanical \& Aerospace Engineering
and LASSP, Cornell University, Ithaca, NY 14853, USA}
\begin{abstract}
We investigate the
non-equilibrium
dynamics of an isolated polymer in a stationary elongational flow.
We compute the relaxation time to the 
steady-state configuration as a function of
the Weissenberg number. A strong increase of the relaxation time is
found around the coil--stretch transition, which is attributed to the  large 
number of polymer configurations. The relaxation dynamics of the 
polymer is solved analytically in terms of a 
central two-point connection problem for the singly confluent Heun equation.
\end{abstract}
\maketitle

\centerline{\textit{J. Phys. A: Math. Gen.} \textbf{39}, 10691--10701 (2006)}
\centerline{\fontfamily{cmr}\selectfont
\tt{statcks.iop.org/JPhysA/39/10691}}

\section{Introduction}
\label{sec:introduction}
Dilute polymer solutions exhibit physical behaviours that
distinguish them from ordinary Newtonian fluids. Even a small
polymer concentration can considerably change the 
large-scale behaviour of the  flow by enhancing viscosity and
reducing the turbulent drag. A comprehensive understanding
of hydrodynamical properties of polymer solutions is still 
lacking despite the large number of
industrial applications (e.g. Sreenivasan and White 2000).
A starting point for the theoretical description  of dilute polymer 
solutions is the dynamics of isolated polymer molecules. The knowledge of how a single molecule
is deformed by the velocity gradient allows the development of  
constitutive models that in turn can be used to calculate the large scale flow.
The stationary dynamics of a single, isolated polymer
has received much attention to date (see Larson 2005, Shaqfeh 2005, for a review). 
On the contrary, less is known about non-equilibrium dynamics of isolated polymers in flow.
Since Rouse (1953) and Zimm's (1956)  seminal works,
experimental, numerical, and theoretical studies focused on the internal 
relaxation dynamics of a polymer floating in a solvent 
under the influence of Brownian fluctuations. 
The two model situations  considered were:
(a)~a polymer suspended in solution and pulled at 
the ends (Quake \textit{et al.} 1997, Hatfield and Quake 1999);
(b)~a tethered polymer submitted to a uniform flow and freely relaxing 
after cessation of the flow (Perkins \textit{et al.} 1994, Brochard-Wyart 1995,
Manneville \textit{et al.} 1996, Rzehak and Zimmermann 2002).
The above studies did not take into account the interaction 
between the polymer and an external flow. Hern\'andez Cifre and Garc\'{\i}a de la Torre (1999)
note that Rouse and Zimm's theories 
may not provide the adequate time scale for coil--stretch processes
in flow on the basis of Brownian Dynamics simulations. 
Here, we investigate how an elongational flow influences
polymer relaxation dynamics. We determine the time scale associated with polymer deformation 
\textit{in the flow} and show a significant deviation from Rouse and Zimm's predictions in the vicinity 
of the coil--stretch transition. 
Our analysis is based on an eigenvalue problem for the confluent Heun equation,
and constitutes a new physical application of this latter equation.

Polymer dynamics is extremely rich already in simple deterministic flows, 
such as elongational flows~(Perkins \textit{et al.} 1997, Smith and Chu 1998),
shear flows~(Smith \textit{et al.} 1999, Celani \textit{et al.} 2005b, Puliafito and Turitsyn 2005,
Gerashchenko and Steinberg 2006) 
or combinations of the two~(Hur \textit{et al.} 2002, Babcock \textit{et al.} 2003).
The velocity gradient of a non-uniform flow stretches the polymer, while  
entropic forces attempt 
to restore the polymer into the coiled equilibrium shape.
In other words, the configuration of the polymer results from the 
counterbalance between 
the entropic forces and the hydrodynamic drag. 
Here we consider an elongational flow, which
 is defined by a constant velocity gradient~$\lambda$.
This flow is particularly effective in stretching polymers
far from their equilibrium configuration. 
A transition to the stretched state of the polymer occurs as 
the velocity gradient exceeds the critical 
value~$\lambda_{c}=1/(2\tau)$, where~$\tau$ 
is the time associated with the slowest relaxation mode of the polymer
in thermal equilibrium with the surrounding medium.
For~$\lambda<\lambda_{c}$ 
polymers stay in the coiled equilibrium configuration;
for~$\lambda>\lambda_{c}$ they become fully extended.
This phenomenon is known as the coil--stretch transition~(de Gennes 1974).
Rouse (1953) computed~$\tau$ as a function of the number of monomers 
in the polymer assuming the polymer could be described as a beads-and-springs
chain with Hookean interactions. Zimm (1956) refined Rouse's prediction by
taking into account hydrodynamic interactions between segments of the chain.

The study of polymer dynamics has benefited from a class of mesoscopic models
that are based on a coarse-grained treatment of the polymer molecule. One 
of the simplest is 
the elastic dumbbell model, which only takes into account the slowest oscillation mode of the molecule (Bird \textit{et al.} 1987).
Notwithstanding this crude simplification, the dumbbell model captures the main
aspects of polymer dynamics in elongational flows, such as the
coil--stretch transition (Perkins \textit{et al.} 1997)
and finite-time conformation hysteresis (Schroeder \textit{et al.} 2003, Schroeder 
\textit{et al.} 2004).
We examine how the probability density function (PDF) of the
extension of a dumbbell-like
molecule approaches its stationary form. We compute 
the typical time it takes for the initial PDF to attain its steady-state form
and show that, in the proximity of the
coil--stretch transition, this time is exceedingly long compared to~$\tau$.
A similar behaviour is encountered in 
white-in-time isotropic random 
flows (Celani \textit{et al.} 2005a, Martins Afonso and Vincenzi 2005); 
however, for an elongational flow 
the amplification of the transient relaxation time is much stronger.

The problem is solved within the framework of the Fokker--Planck equation 
for the PDF of the extension of the polymer.
The computation of the transient relaxation time is recast as a central 
two-point connection problem~(CTCP) for a singly confluent Heun 
equation.
The Heun equation is the general Fuchsian differential equation with
four regular singularities.
Its singly confluent form, also known as the generalised 
spheroidal wave equation,  results from the merging of 
two regular singularities into one irregular singularity of
Poincar\'e rank~$1$
(see Slavyanov 1995, for a review). A CTCP is an eigenvalue problem 
for an ordinary differential equation where: (a)~at 
both endpoints of the
interval of definition is located a singularity of the equation;
(b)~the solutions are required to have a specified asymptotic behaviour
while approaching the two singularities from inside.
Contrary to the hypergeometric equation, 
an explicit formula for the CTCP is not known for the Heun family
of equations. The eigenvalues can be determined only as solutions of 
transcendental equations involving continued fractions.
Well-known applications of the confluent Heun equation 
are the electronic spectra of
the hydrogen-molecule ion in the
Born-Oppenheimer approximation (Jaff\'e 1934) and
the Teukolsky equation describing small perturbations to the Kerr geometry
in black hole theory (Leaver 1985).
For a comprehensive introduction to physical applications of the
Heun equations we refer the reader to the book by 
Slavyanov and Lay (2000). 

The rest of this paper is organised as follows. The 
dumbbell model is introduced in 
section~\ref{sec:dumbbell}. The computation of the
transient relaxation time and the related CTCP for the confluent
Heun equation are presented 
in section~\ref{sec:stationary}. 
Section~\ref{sec:conclusions} is devoted to conclusions.

\section{Elastic dumbbell model}
\label{sec:dumbbell}
When one is interested in the statistics of polymer extension
the dynamics of a polymer mole\-cule can be described, 
to a first approximation, in terms of its slowest oscillation mode.
In this case, the molecule can be modelled as an elastic dumbbell,
i.e. as two beads connected by a
spring.
The beads represent the ends of the molecule and
the spring models entropic forces. The separation between the beads
measures the extension of the polymer.
When introduced into a 
non-uniform flow the molecule experiences collisions with 
fluid particles and becomes stretched under the action of the velocity 
gradient. In the simplest case, the drag force 
is assumed to be proportional to the velocity of the polymer relative 
to the fluid and thermal agitation is modelled by Brownian motion.
In most  applications the extension of the polymer remains smaller than the 
dissipative scale of the carrier flow, and therefore the velocity field can be 
assumed to be linear%
\footnote{Davoudi and Schumacher (2006) recently investigated the situation
where the extension of the polymer can reach the inertial range of 
turbulence.}.
The separation vector of the ends of the polymer, $\bm{Q}$,
is then a stochastic process evolving according to
the three-dimensional stochastic ordinary differential equation%
\footnote{There is no Ito--Stratonovich ambiguity in this case
because the coefficient of Brownian motion does not depend 
on $\bm{Q}$.} (e.g. Bird \textit{et al.} 1987)
\begin{equation}
\label{eq:dumbbell}
d\bm{Q}=\bm{Q}\cdot\nabla\bm{v}(s)\, ds-
f(Q)\,\frac{\bm{Q}}{2\tau}\, ds
+\sqrt{\frac{Q_0^2}{\tau}}\, d\bm{B}(s),
\end{equation}
where $Q=|\bm{Q}|$, $Q_0^2$ is the equilibrium mean-square separation 
of the ends of the molecule, and~$\bm{B}(s)$ 
denotes the three-dimensional standard Brownian motion.
The function~$f$ determines the entropic force. We consider the
finitely extensible nonlinear elastic (FENE) dumbbell model with 
\begin{displaymath}
f(Q)=\frac{1}{1-Q^2/L^2}, \qquad 
0\le Q< L,
\end{displaymath}
where~$L$ is the maximum extension of the molecule.
The force diverges as the molecular extension
approaches~$L$; thus the extension of the molecule will be finite and smaller than~$L$.
The FENE model is appropriate for synthetic
polymers such as polyacrylamide and polyethyleneoxide;
biological macromolecules such as DNA and polyeptides
are better described by the worm-like chain model (e.g. Larson 2005).
In elongational-flow experiments, the flexibility parameter
$b=L^2/Q_0^2$ usually ranges from~$10^2$ to~$10^4$
(Larson 2005, Shaqfeh 2005). 

The FENE model~\eqref{eq:dumbbell}
 neglects hydrodynamic interactions between the segments of the 
polymer. Inclusion of hydrodynamic interactions makes the coil--stretch
transition sharper, but do not change  the transition qualitatively
(e.g. Wiest \textit{et al.} 1989, Hern\'andez Cifre and Garc\'{\i}a de la Torre 1998, 1999).
Furthermore,
the dumbbell model without hydrodynamic interactions accurately reproduces 
the extension--strain curve observed in experiments (Perkins \textit{et al.} 1997).
The accuracy of the model can be ascribed to the
cancellation between the effect of the 
distribution of drag forces along the chain
and the effect of the increase in effective drag coefficient with polymer 
extension (Larson \textit{et al.} 1997). Neglecting hydrodynamic interactions
has the benefits of providing analytical results.

\section{Relaxation dynamics}
\label{sec:stationary}
A steady 
planar elongational flow is characterised by one direction of stretch, one direction of 
compression, and one neutral direction. The velocity gradient is 
constant along the directions both of the extensional and compressional axes:
$\nabla_jv_i=\lambda(\delta_{i1}\delta_{j1}-\delta_{i2}\delta_{j2})$,
$\lambda>0$.
When a polymer is immersed in that flow the 
first component of the separation vector rapidly becomes much greater than the 
other components: $Q_1\gg Q_2$ and~$Q_1\gg Q_3$. 
The extension of the molecule, therefore, is approximatively 
$Q\simeq Q_1$, and $f(Q)$ can be replaced by 
$f(Q_1)$.\footnote{%
The above approximation holds true for a steady uniaxial extensional flow as well:
$\nabla_jv_i=\lambda(\delta_{i1}\delta_{j1}-\delta_{i2}\delta_{j2}/2
-\delta_{i3}\delta_{j3}/2)$.
Therefore, the results henceforth presented are 
unchanged for this flow.}
According to this approximation, the first component of~\eqref{eq:dumbbell}
gives a one-dimensional stochastic differential
equation for the extension of the dumbbell:
\begin{equation}
\label{eq:extension}
dQ=\lambda Q\,d s-
f(Q)\,\frac{Q}{2\tau}\, d s
+\sqrt{\frac{Q_0^2}{\tau}}\, dB(s),
\end{equation}
where $B(s)$ is the one-dimensional Brownian motion.
For the sake of notational simplicity, we introduce the rescaled 
separation vector~$\bm{q}=\bm{Q}/L$. The PDF of~$q=\vert\bm{q}\vert$, $\psi(q;t)$ with
$q\in[0,1]$, satisfies the one-dimensional Fokker--Planck equation associated 
with~\eqref{eq:extension} (e.g. Stratonovich 1963)
\begin{equation}
\label{eq:FPE}
\begin{array}{l}
\partial_t \psi=\mathscr{L}\psi,\\
\displaystyle
\mathscr{L}\psi:=-\frac{\partial}{\partial q}
\bigg[\bigg(\mathit{Wi}-\frac{\hat{f}(q)}{2}\bigg)q\psi\bigg]+
\frac{1}{2b}\,\frac{\partial^2 \psi}{\partial q^2},
\end{array}
\end{equation}
where $t=s/\tau$, $\hat{f}(q)=f(Lq)$, and $\textit{Wi}=\lambda\tau$.
The dimensionless number~\textit{Wi} is known as the Weissenberg number and
measures the level of polymer stretching. For~$\textit{Wi}<1/2$
polymers are in the coiled state; for~$\textit{Wi}>1/2$ polymers are fully 
extended. The critical value 
$\textit{Wi}_{c}=\lambda_{c}\tau=1/2$ marks the 
coil--stretch transition in elongational flows (de Gennes 1974).
Strictly speaking, the approximation leading
to~\eqref{eq:extension} and~\eqref{eq:FPE} holds true when the polymer is
sufficiently stretched, i.e. only for~$\textit{Wi}\gtrsim\textit{Wi}_c$.
Equation~\eqref{eq:FPE}, indeed, does not yield a good approximation of~$\psi(q;t)$
at low~$\mathit{Wi}$.
However, we shall see at the end of this section that
it is appropriate to investigate
polymer relaxation dynamics in terms of~\eqref{eq:extension} and~\eqref{eq:FPE}
also below the coil-stretch transition.

The Fokker-Planck equation~\eqref{eq:FPE} 
is solved with reflecting boundary conditions, that is 
the probability current
\begin{displaymath}
j(q;t)=\textit{Wi}\: q\psi(q;t)-\frac{\hat{f}(q)}{2}q\psi(q;t)-\frac{1}{2b}
\left.\frac{\partial \psi}{\partial q}\right|_{q,t}
\end{displaymath}
is assumed to vanish at the endpoints of the interval of definition:
$j(0;t)=j(1;t)=0$ $\forall\,t>0$ (Stratonovich 1963). 
Under these conditions,
the stationary distribution of the extension, $\psi_0(q)=\lim_{t\to\infty}
\psi(q;t)$, can be derived by simple integration (Bird \textit{et al.} 1987):
\begin{equation}
\label{eq:stationary}
\psi_0(q)=N e^{b\mathit{Wi}\, q^2}(1-q^2)^\frac{b}{2},
\qquad 0\leq q\leq 1,
\end{equation}
where
\begin{displaymath}
N=2\Gamma((b+3)/2)\,/\,[\sqrt{\pi}\,
\Gamma(b/2+1)\,\times\, _1F_1(1/2;(b+3)/2;b\textit{Wi})]
\end{displaymath}
and~$\Gamma$ and~$_1F_1$ denote the Euler Gamma function and the 
confluent hypergeometric function, respectively (e.g. Erd\'elyi \textit{et al.} 1953).
Large extensions become more and more probable with 
increasing~\textit{Wi} in accordance with experiments (Perkins \textit{et al.} 1997).
Note however that, at low~\textit{Wi},
equation~\eqref{eq:stationary} does not constitutes a
good description of the PDF of small extensions due to the 
approximation behind~\eqref{eq:FPE}.

The time behaviour of the system depends on the form of the spectrum of the
operator~$\mathscr{L}$ with reflecting boundary conditions:
\begin{eqnarray}
\label{eq:spectrum}
\mathscr{L}\psi_\mu=-\mu \psi_\mu,\\
\label{eq:bc}
j_\mu(0)=j_\mu(1)=0,
\end{eqnarray}
where $j_\mu$ denotes the probability
current associated with the eigenfunction~$\psi_\mu$.
Under conditions~\eqref{eq:bc}, $\mathscr{L}$
is non-positive defined and symmetric with respect to the scalar product 
with weighting function~$1/\psi_0$. Its eigenvalues~$\mu$ are therefore
real and non-negative, $\mu=0$
being associated with the long-time solution~$\psi_0$ (Stratonovich 1963). 
If~$\mathscr{L}$ has a countable spectrum~$\{0,\mu_1,\mu_2,\dots\}$
with~$\mu_i<\mu_{i+1}$, then~\mbox{$T_{\mathrm{rel}}=\tau/\mu_1$}
is the characteristic time needed for~$\psi(q;t)$
to attain its long-time form~$\psi_0(q)$ when the initial 
condition~$\psi(q;0)$ is taken far from equilibrium (Schenzle and Brand 1979). 
We note that while~$\tau$ is the relaxation time
of the polymer \textit{in the absence 
of flow}, $T_{\mathrm{rel}}$ characterises polymer relaxation 
dynamics \textit{in the flow}. 

To compute~$T_{\mathrm{rel}}$, we need to solve the eigenvalue 
problem~\eqref{eq:spectrum}, \eqref{eq:bc}.
By making the substitutions~$z=q^2$, 
$\psi_\mu(z)=(1-z)^{\frac{b}{2}}w_\mu(z)$, we can 
rewrite~(\ref{eq:spectrum}) in the form
\begin{equation}
\label{eq:CHE}
w''_\mu+\bigg(\beta+\frac{\gamma}{z}+\frac{\delta}{z-1}\bigg)w'_\mu
+\frac{\alpha\beta z-\nu}{z(z-1)}\,w_\mu=0,
\end{equation}
where $w'_\mu:=dw_\mu/dz$, $\gamma=1/2$, $\delta=b/2$, 
$\alpha=(1+b-\mu/\mathit{Wi})/2$, $\beta=-b\mathit{Wi}$, and 
$\nu=b(\mu-\mathit{Wi})/2$.
The above equation is a singly confluent Heun equation in 
the non-symmetrical canonical form
(Decarreau \textit{et al.} 1978a, Decarreau \textit{et al.} 1978b, Slavyanov 1995)%
\footnote{%
The worm-like chain model, $\hat{f}(q)=2/3-1/(6q)+1/[6q(1-q)^2]$,
would lead to a second-order linear differential equation with two
irregular singularities of Poincar\'e rank~1 and~3, respectively.
The FENE model in a $\delta$-correlated random flow can be solved in terms of a  general Heun equation
(Martins Afonso and Vincenzi 2005).}.

Equation~(\ref{eq:CHE}) has two regular singularities at~$z=0$ and $z=1$
and an irregular singularity of Poincar\'e rank 1 at~$z=\infty$.
The characteristic exponents 
at~$z=0$ are~0 and~$1-\gamma$; at~$z=1$, they are~0 and~$1-\delta$.
We neglect the case of~$b$ integer and odd since it does not have
physical relevance.
Consequently, there are no logarithmic solutions and,
for $|z|<1$, $w_\mu(z)$ can be written as $w_\mu(z)
=a_0\varphi_0(z)+d_0z^{1/2}\chi_0(z)$ with $a_0$, $d_0$ complex constants and
$\varphi_0$, $\chi_0$ analytic functions of~$z$ (equivalently of~$q^2$) 
such that~$\varphi_0(0)\ne 0$ and~$\chi_0(0)\ne 0$.
Similarly,
for $|z-1|<1$, $w_\mu(z)$ takes the form~$w_\mu(z)=a_1\varphi_1(z-1)+d_1
(1-z)^{1-b/2}\chi_1(z-1)$ with $a_1$, $d_1$ complex constants and $\varphi_1$,
$\chi_1$ analytic functions of~$z$ such 
that~$\varphi_1(0)\neq 0$ and~$\chi_1(0)\neq 0$.

It is easily seen that boundary conditions~(\ref{eq:bc}) 
can be matched only by those solutions
of~(\ref{eq:CHE}) that belong to the exponent~0 both 
in $z=0$ and $z=1$ (that is both $d_0$ and $d_1$ have to be zero)\footnote{
We leave aside the case $b=2$
since it does not have physical relevance.}. The eigenvalue problem defined 
by~(\ref{eq:spectrum}) and (\ref{eq:bc}) is therefore mapped into a
CTCP  on $[0,1]$ for the confluent Heun equation~(\ref{eq:CHE}).
To compute the eigenvalues~$\mu$, we adapt the procedure exploited
by Svartholm (1939) and Erd\'elyi (1942, 1944)
to~(\ref{eq:CHE}) (see also Slavyanov 1995, Slavyanov and Lay 2000).
We then expand~$w_\mu$ in series of
Jacobi polynomials having the required characteristic exponent 
at~$z=0$ and~$z=1$:
\begin{equation}
\label{eq:expansion}
w_\mu(z)=\sum_{n=0}^{\infty}c_n(\mu) u_n(z)
\end{equation}
with
\begin{displaymath}
u_n(z)=\,
_2F_1(-n,\omega+n;\gamma;z)=(-1)^n\frac{n!\sqrt{\pi}}{\Gamma(n+\gamma)}
\,P_n^{(\omega-\gamma,\gamma-1)}(2z-1)
\end{displaymath}
and $\omega=\gamma+\delta-1$. 
The function~$_2F_1$ denotes the Gauss hypergeometric function; 
$P_n^{(\omega-\gamma,\gamma-1)}$ are the Jacobi polynomials of 
parameters~$\omega-\gamma$ and~$\gamma-1$ and degree~$n$
(e.g. Erd\'elyi \textit{et al.} 1953).
The conditions for the convergence 
of~\eqref{eq:expansion} will determine the spectrum of~$\mathscr{L}$.

By defining the operators
\begin{eqnarray*}
\mathscr D_1w_\mu:=z(z-1)
\bigg[w''_\mu+\bigg(\frac{\gamma}{z}+\frac{\delta}{z-1}\bigg)w'_\mu\bigg],\\
\mathscr D_2w_\mu:=-z(z-1)w'_\mu,
\end{eqnarray*}
we can rewrite~(\ref{eq:CHE}) in the form
\begin{equation}
\label{eq:CHE_2}
\mathscr D_1w_\mu+\varepsilon \mathscr D_2w_\mu+(\rho z+\sigma)w_\mu=0,
\end{equation}
where $\varepsilon=b\mathit{Wi}$, $\rho=b[\mu-(1+b)\mathit{Wi}]/2$, 
$\sigma=b(\mathit{Wi}-\mu)/2$. The polynomials~$u_n$ satisfy the
differential relations (Erd\'elyi \textit{et al.} 1953)
\begin{equation}
\label{eq:pol_1}
\mathscr{D}_1u_n=n\left(\omega+n\right)u_n
\end{equation}
and
\begin{equation}
\label{eq:pol_2}
\mathscr{D}_2u_n=\widetilde{A}_nu_{n+1}+\widetilde{B}_nu_{n}+
\widetilde{C}_nu_{n-1}
\end{equation}
with
\begin{eqnarray*}
\widetilde{A}_n=\frac{n(n+\omega)(n+\gamma)}{(2n+\omega)(2n+\omega+1)},
\quad
\widetilde{B}_n=\frac{n(n+\omega)(\delta-\gamma)}{(2n+\omega-1)(2n+\omega+1)},
\\[0.2cm]
\widetilde{C}_n=-\frac{n(n+\omega)(n+\delta-1)}{(2n+\omega)(2n+\omega-1)}\,.
\end{eqnarray*}
In addition, the following recurrence relation holds (Erd\'elyi \textit{et al.} 1953):
\begin{equation}
\label{eq:pol_3}
zu_n=\widehat{A}_nu_{n+1}+\widehat{B}_nu_n+\widehat{C}_nu_{n-1}
\end{equation}
with coefficients
\begin{eqnarray*}
\widehat{A}_n=-\frac{(n+\gamma)(n+\omega)}{(2n+\omega)(2n+\omega+1)},
\quad
\widehat{B}_n=\frac{2n(n+\omega)+\gamma(\omega-1)}{(2n+\omega-1)(2n+\omega+1)},\\[0.2cm]
\widehat{C}_n=-\frac{n(n+\delta-1)}{(2n+\omega-1)(2n+\omega)}\,.
\end{eqnarray*}
Inserting expansion~(\ref{eq:expansion}) in~(\ref{eq:CHE_2}) and
exploiting relations~(\ref{eq:pol_1}), (\ref{eq:pol_2}), (\ref{eq:pol_3})
yield the following three-term recurrence relation for~$c_n(\mu)$:
\begin{eqnarray}
\label{eq:iniz_1}
g_{-1}c_0=0&\\
\label{eq:iniz_2}
g_0c_1+h_0c_0=0&\\
\label{eq:3}
g_nc_{n+1}+h_nc_n+k_nc_{n-1}=0& \quad \mathrm{for}\ n\ge 2,
\end{eqnarray}
where the coefficients
\begin{eqnarray*}
g_n=\varepsilon\widetilde{C}_{n+1}+\rho\widehat{C}_{n+1},\qquad
h_n=n(\omega+n)+\varepsilon\widetilde{B}_n+\rho\widehat{B}_n+\sigma,
\\
k_n=\varepsilon\widetilde{A}_{n-1}+\rho\widehat{A}_{n-1}
\end{eqnarray*}
depend on~$\mu$ through~$\rho$ and~$\sigma$.
The asymptotic behaviour of~$h_n/g_n$ and~$k_n/g_n$ is given by
\begin{equation}
\label{eq:asympt}
\frac{h_n}{g_n}\sim
-\frac{4n}{\varepsilon}\qquad\mathrm{and}\qquad
\frac{k_n}{g_n}\sim -1\qquad (n\to\infty).
\end{equation}
According to the Perron-Kreuser theorem (e.g. Gautschi 1967, Wimp 1984), 
\eqref{eq:asympt}
implies that the recurrence rela\-tion~(\ref{eq:3}) has two linearly independent 
solutions~$(c_{n}^+)_n$ and~$(c_{n}^-)_n$ such that
\begin{equation}\label{eq:minimal}
\frac{c^+_{n+1}}{c^+_{n}}\sim\frac{4n}{\varepsilon}
\qquad \mathrm{and} \qquad
\frac{c^-_{n+1}}{c^-_{n}}\sim-\frac{\varepsilon}{4n}
\qquad (n\to\infty).
\end{equation}
The above sequences have the property: $\lim_{n\to\infty}c^-_n/c^+_n=0$;
hence $(c^-_n)_n$ is what is called a minimal solution of~\eqref{eq:3}.
Every other solution non proportional to~$(c^-_n)_n$
is asymptotically proportional to~$(c^+_n)_n$, and 
therefore diverges with increasing~$n$. Therefore, we can already say 
that~\eqref{eq:expansion} can converge only if the 
sequence~$(c_n)_n$ is a minimal solution of~\eqref{eq:3}.  

The second limit in~\eqref{eq:minimal} implies that
$\lim_{n\to\infty}\sqrt[n]{|c^-_n|}=0$. This condition ensures 
that, if the coefficients~$c_n$ form a minimal solution of~\eqref{eq:3}, then 
expansion~\eqref{eq:expansion}
convergences absolutely  to an analytic 
function in the whole complex plane (Szeg\"o 1939, p.~252).

Nonetheless,
it is also required that~$c_n=0$ for~$n<0$, that is the 
sequence~$(c_n)_n$ is subject  to initial conditions~\eqref{eq:iniz_1} 
and~\eqref{eq:iniz_2}.
Equation~\eqref{eq:iniz_1} is trivially satisfied since~$g_{-1}=0$, and so 
does~\eqref{eq:iniz_2} for~$\mu=0$. (This is in accordance with the fact 
that~$\psi_0$ exists for all~$\textit{Wi}$ and~$b$.)
However, for~$\mu\ne 0$, equation~\eqref{eq:iniz_2} fixes the 
ratio~$c_1/c_0=-(b+3)/2$. This latter requirement is not satisfied by 
whatever~$\mu$, and therefore select the spectrum of~$\mathscr{L}$.

To summarise, the spectrum of~$\mathscr{L}$ is the set 
of those~$\mu$ such that~$(c_n(\mu))_{n\ge 0}$ is a minimal
solution of~\eqref{eq:3} satisfying~$c_1(\mu)/c_0(\mu)=-(b+3)/2$. 
From Pincherle's theorem (e.g. Gautschi 1967, p.~31), this conclusion is equivalent
to stating that the eigenvalues~$\mu$ are the solutions of
\[\displaystyle
\frac{k_1/g_1}{h_1/g_1-
\displaystyle\frac{k_2/g_2}{h_2/g_2-\displaystyle\frac{k_3/g_3}{h_3/g_3-
\dots}}}
=\frac{b+3}{2}.
\]
The latter represents a transcendental equation 
for the relaxation spectrum associated with the Fokker-Planck 
equation~\eqref{eq:FPE}. The reciprocal of the smallest
nonzero solution is~$T_{\mathrm{rel}}/\tau$.

\begin{figure}
\begin{center}
\includegraphics[width=0.6\textwidth]{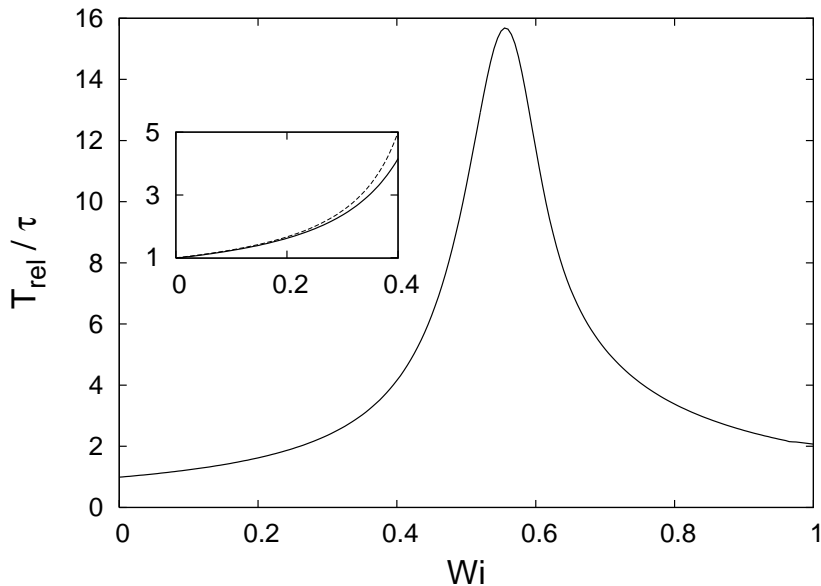}
\end{center}
\caption{\label{fig}Rescaled relaxation time~$T_{\mathrm{rel}}/\tau$ 
as a function of the Weissenberg
number~\textit{Wi} for~$b=600.25$; the inset shows the comparison
with the small-\textit{Wi} behaviour $1/(1-2\mathit{Wi})$ (dashed line).
}
\end{figure}

\begin{figure}
\begin{center}
\includegraphics[width=0.325\textwidth]{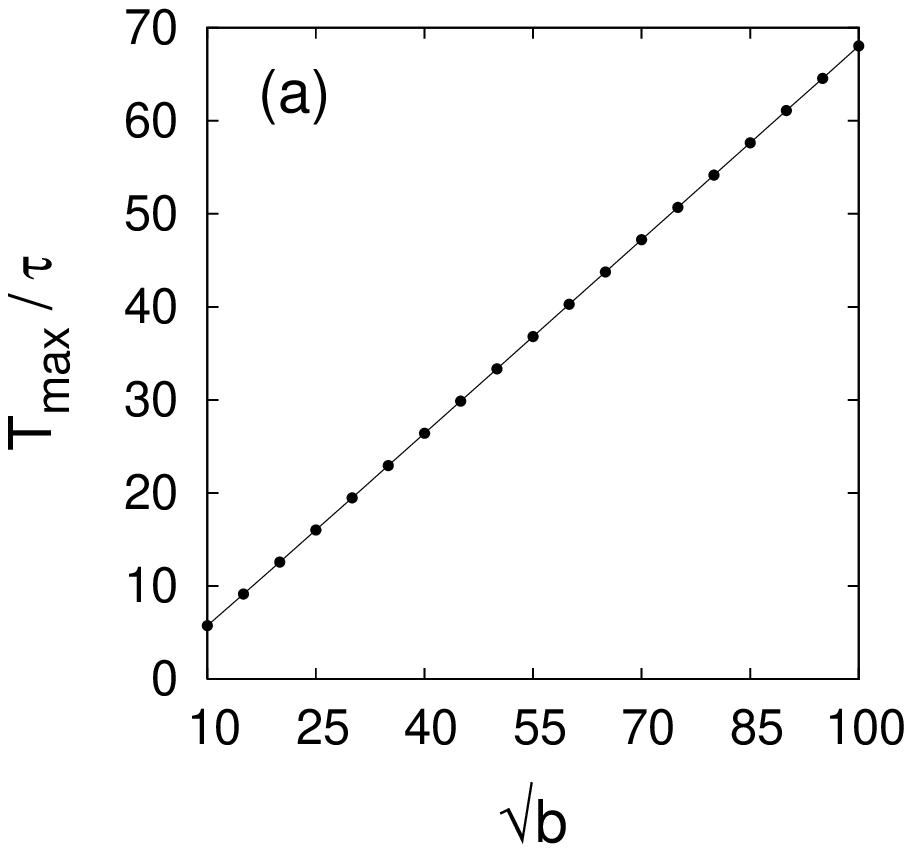}\hspace{0.2cm}
\includegraphics[width=0.35\textwidth]{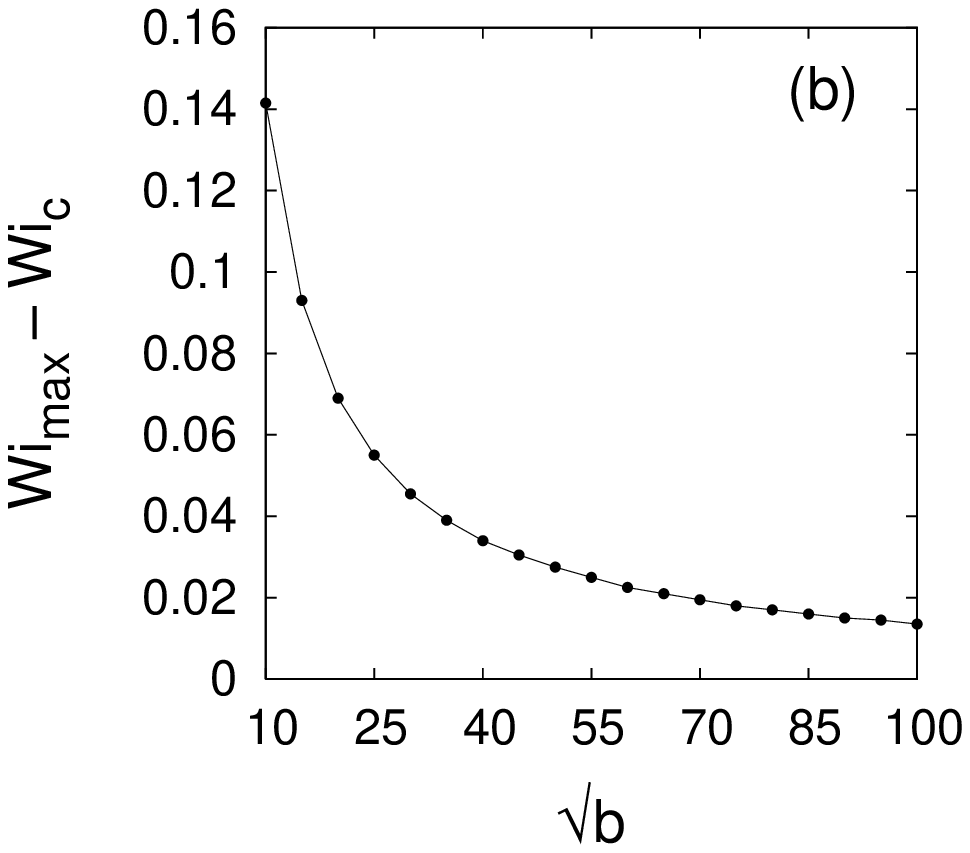}\\[0.4cm]
\includegraphics[width=0.62\textwidth]{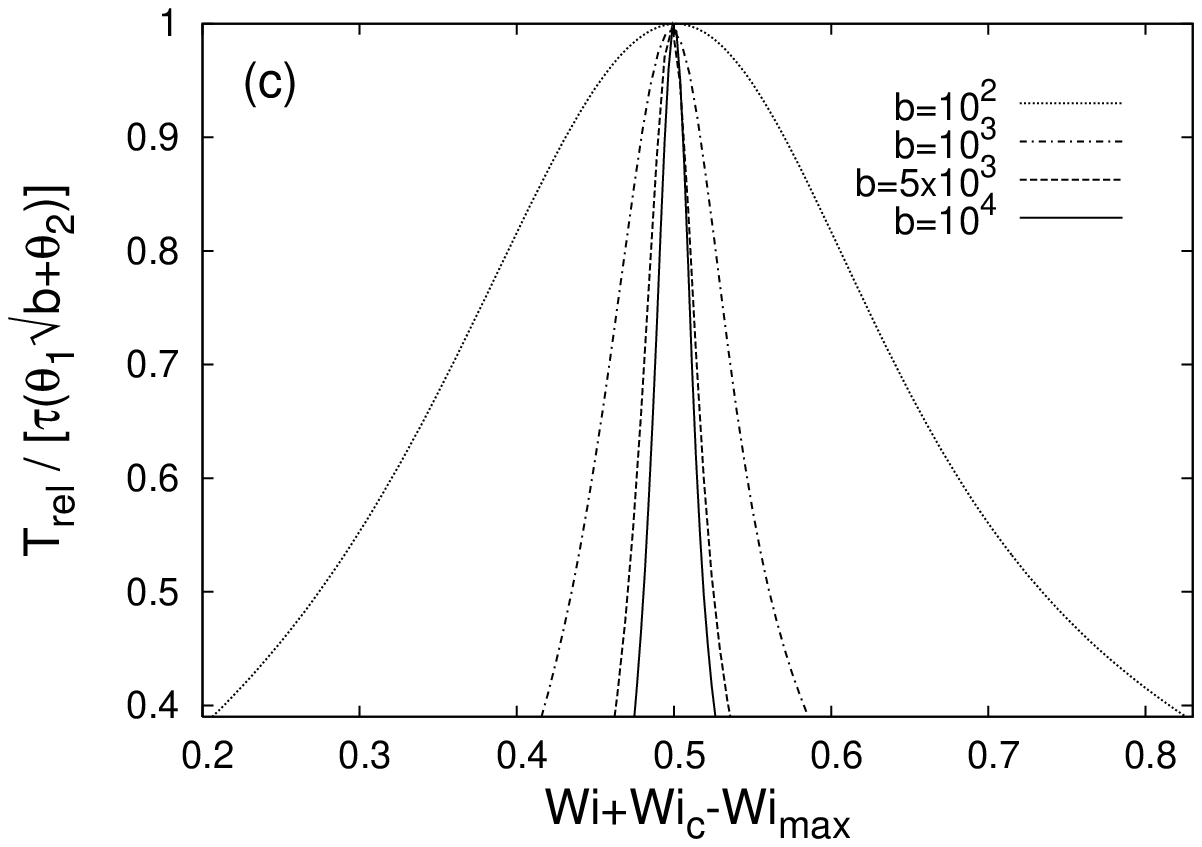}
\end{center}
\caption{(\textit{a}) Maximum rescaled relaxation 
time~$T_{\mathrm{max}}/\tau$ vs. the square root
of the flexibility parameter~$b=L^2/Q_0^2$: $T_{\mathrm{max}}/\tau$
increases as $\theta_1\sqrt{b}+\theta_2$
with $\theta_1\approx 0.69$ and $\theta_2\approx -1.18$; (\textit{b}) 
Behaviour of~$\mathit{Wi}_{\mathrm{max}}$ as a function of~$\sqrt{b}$, 
where~$\mathit{Wi}_{\mathrm{max}}$ is the Weissenberg
number at which the relaxation time attains its maximum 
value~$T_{\mathrm{max}}$;
(\textit{c})~Relaxation time vs.~\textit{Wi} for different~$b$,
rescaled to make the maximum values coincide.}
\label{fig:2}
\end{figure}

We calculated~$T_{\mathrm{rel}}$ numerically by means of
the modified Miller algorithm (Wimp 1984, pp.~82--85).
The behaviour of~$T_{\mathrm{rel}}$ as a function of~\textit{Wi}
is reported in fig.~\ref{fig}.
For very small~$\textit{Wi}$ the influence of the external flow is negligible
and~$T_{\mathrm{rel}}$ is approximately equal to~$\tau$.
With increasing~$\textit{Wi}$ the ratio~$T_{\mathrm{rel}}/\tau$ starts 
growing as~$1/(1-2\textit{Wi})$.
This behaviour can be deduced by 
replacing the spring force with a Hookean force, $\hat{f}(q)=1$, 
and computing the relaxation of the moments of~$q$.
For a fixed~$b$, the rescaled relaxation time reaches a
sharp maximum~$T_{\mathrm{max}}$ in the neighbourhood of the coil--stretch 
transition ($\textit{Wi}_{c}=1/2$);
the value of~$T_{\mathrm{max}}$ increases 
linearly with~$\sqrt{b}$ (fig.~\ref{fig:2}). With increasing~$b$,
$T_{\mathrm{rel}}$ attains its maximum value~$T_{\mathrm{max}}$ 
closer and closer 
to~$\textit{Wi}_{c}$ and the width of the peak decreases
(fig.~\ref{fig:2}).
For large~$\textit{Wi}$ the relaxation time is fixed by the time scale
of the flow~$\lambda^{-1}$, and therefore~$T_{\mathrm{rel}}/\tau$ decreases 
as~$\textit{Wi}^{-1}$.

The computation of~$T_{\mathrm{rel}}$ shows that,
near the coil--stretch transition, the 
typical time scale involved in polymer relaxation
dynamics is strongly different from~$\tau$. The time associated with the
fundamental oscillation mode 
is not representative of polymer temporal dynamics in flow.
These results confirm the behaviour 
encountered in isotropic $\delta$-correlated Gaussian flows
(Celani \textit{et al.} 2005a, Martins Afonso and Vincenzi 2005), 
but here the 
enhancement of~$T_{\mathrm{rel}}$ is stronger since the elongational flow
is more efficient in deforming polymers.

\begin{figure}[t]
\includegraphics[width=0.47\textwidth]{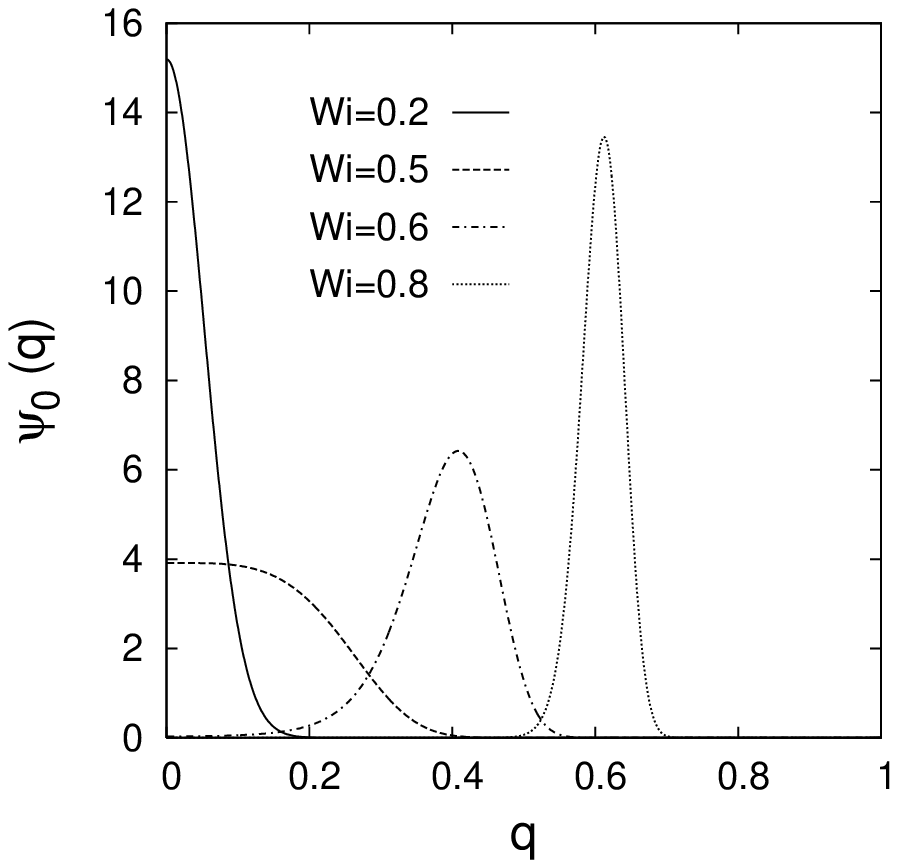}
\hfill
\includegraphics[width=0.458\textwidth]{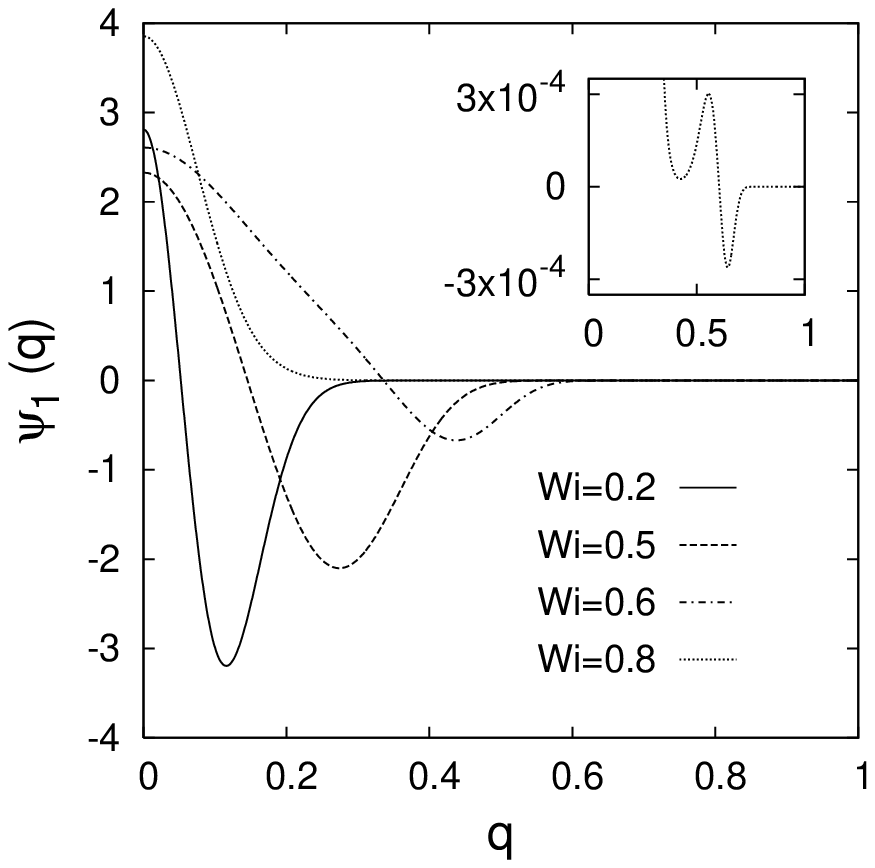}
\caption{Stationary PDF of the extension and eigenfunction of the Fokker--Planck operator~$\mathscr{L}$ 
associated with the smallest nonzero eigenvalue~$\mu_1$ ($b=600.25$).
The eigenfunctions have been computed numerically by means of the iteration--variation method 
(Morse and Feshbach 1953).}
\label{fig:eigen}
\end{figure}

In Brownian Dynamics simulations and experiments the relaxation time of~$\psi(q;t)$ can be measured
from the decay of the moments of the extension to their steady-state value. 
To observe~$T_{\mathrm{rel}}$, 
one has to make sure that the initial condition~$\psi(q;0)$ is not orthogonal to the first 
eigenfunction of~$\mathscr{L}$, that is $\psi_1(q)$ in~\eqref{eq:spectrum}, with respect to the scalar
product with weighting function~$\psi_0(q)$. 
Figure~\ref{fig:eigen} shows the 
shape of~$\psi_1(q)$ for different~$\textit{Wi}$. 
An initial PDF concentrated either at a coiled or a stretched extension~$\bar{q}$
(formally $\psi(q;0)=\delta(q-\bar{q})$) has projection onto~$\psi_1(q)$ equal 
to~$\psi_1(\bar{q})/\psi_0(\bar{q})$ by construction (e.g. Stratonovich 1963).
The initial condition with all the polymers having the same extension~$\bar{q}$ is therefore not orthogonal
to~$\psi_1(q)$ if~$\psi_1(\bar{q})\neq 0$; 
all the moments will thus tend to their stationary value with a typical time scale~$T_{\mathrm{rel}}$.

We now show that it is accurate to compute~$T_{\mathrm{rel}}$ by means of~\eqref{eq:FPE} 
also for~$\textit{Wi}<\textit{Wi}_c$. 
At low~$\textit{Wi}$, the entropic force can be modelled as
a Hookean force, i.e.~$f(Q)=1$ in~\eqref{eq:dumbbell}.
The vector equation~\eqref{eq:dumbbell} therefore reduces to a set of three decoupled stochastic
differential equations for $Q_1$, $Q_2$, $Q_3$. Hence we 
have~$\Psi(\bm{q};t)=\psi^{(1)}(q_1;t)\psi^{(2)}(q_2;t)\psi^{(3)}(q_3;t)$, 
where~$\Psi(\bm{q};t)$ is the PDF of the vector~$\bm{q}$ 
and~$\psi^{(i)}(q_i;t)$ is the PDF of the $i$-th component of~$\bm{q}$.
The relaxation time of~$\Psi(\bm{q};t)$ is thus the longest of the 
relaxation times of~$\psi^{(1)}(q_1;t)$, $\psi^{(2)}(q_2;t)$,
and~$\psi^{(3)}(q_3;t)$ given that all the~$\psi^{(i)}(q_i;t)$ have a non-trivial
long-time limit.
It is easy to check that at low~\textit{Wi} the relaxation time
of~$\psi^{(1)}(q_1;t)$ is~$T_{\mathrm{rel}}^{(1)}=\tau/(1-2\mathit{Wi})$, 
while~$T_{\mathrm{rel}}^{(2)}=\tau/(1+2\mathit{Wi})$ and~$T_{\mathrm{rel}}^{(3)}=\tau$ (with obvious
notation). The function $\Psi(\bm{q};t)$ therefore has the same relaxation time 
as~$\psi^{(1)}(q_1;t)$.
Now we note that~$\psi(q;t)$ and~$\Psi(\bm{q};t)$ have the same long-time behaviour
since~$\psi(q;t)=q^2\int_0^{\pi}\int_0^{2\pi} \Psi(\bm{q}(q,\vartheta,\phi);t) \sin{\vartheta}\,d\vartheta d\phi$. We thus come to the following conclusion:
the relaxation times of~$\psi(q;t)$ and~$\psi^{(1)}(q_1;t)$
coincide at low~\textit{Wi} and $\psi^{(1)}(q_1;t)$ satisfies~\eqref{eq:FPE}.
This fact explains why~\eqref{eq:FPE} yields an accurate computation of~$T_{\mathrm{rel}}$ also below the
coil--stretch transition.

\section{Summary and conclusions}
\label{sec:conclusions}
We have investigated the relaxation dynamics of an isolated polymer in an 
external flow. Previous studies focused on the relaxation of
a polymer  in a solvent driven only by Brownian
motion. Those studies 
determined the time~$\tau$ associated with the slowest oscillation mode
of the molecule in the absence of external flow.
We have considered an elongational flow and derived a transcendental equation for the relaxation
spectrum associated with the time evolution of the PDF of polymer extension.
The problem of computing this spectrum has been recast as
a central two-point connection problem for a confluent Heun equation. 
The Heun equation results from the separation of the time variable
in the Fokker--Planck equation for the PDF of the extension.

Our analysis shows that an external elongational flow strongly influences
polymer relaxation dynamics.
The longest relaxation time associated with the evolution of the PDF of the
extension provides an estimation of the time scale of polymer deformation. 
Near the coil--stretch transition,
this time is significantly greater than~$\tau$ 
(already one order of magnitude greater
for short molecules). 
The physical reason is the large number of polymer configurations close to the coil--stretch transition. At intermediate
Weissenberg numbers polymer conformation results from a critical 
competition between the entropic force and the velocity gradient. 
Therefore, around the transition,   
coiled and  stretched polymer coexist in the flow, and this makes
the relaxation to the equilibrium PDF particularly long.
It is worth noticing that the amplification of the relaxation
time is stronger when hydrodynamic interactions are taken into account; 
this behaviour is intimately related to the
finite-time conformation hysteresis observed in elongational 
flows (Celani \textit{et al.} 2006).

\acknowledgments
The authors are grateful to A.~Celani, C.~Doering, M.~Martins Afonso, 
A.~Puliafito, and B.~Rajaratnam for useful
suggestions. This work has been partially supported by the European Union
under the contract HPRN-CT-2002-0030 
``Fluid Mechanical Stirring and Mixing: the Lagrangian Approach''.

\begin{itemize}
\addtolength{\itemsep}{-0.25cm}
\item[] Babcock~H~P, Teixeira~R~E, Hur~J~S, Shaqfeh~E~S and Chu~S
2003 {\it Macromolecules} {\bf 36} 4544--8
\item[] Bird R B, Hassager O, Armstrong R C and Curtiss C F 1987
{\it Dynamics of Polymeric Liquids, Fluid Mechanics} vol~2
(New York: Wiley)
\item[] Brochard-Wyart F 1995 \textit{Europhys. Lett.} \textbf{30} 387--92
\item[] Celani A, Musacchio S and Vincenzi D 2005a 
{\it J. Stat. Phys.} {\bf 118} 531--54
\item[] Celani A, Puliafito A and Turitsyn K 2005b {\it Europhys. Lett.} 
{\bf 70} 464--70
\item[] Celani A, Puliafito A and Vincenzi D 2006 \textit{Phys. Rev. Lett.} 
\textbf{97} 118301 
\item[] Davoudi J and Schumacher J 2006 {\it Phys. Fluids} \textbf{18} 025103
\item[] Decarreau~A, Dumont-Lepage~M-Cl, Maroni~P, Robert~A and Ronveaux~A
1978a {\it Ann. Soc. Sci. Bruxelles} {\bf 92} 53--78
\item[] Decarreau A, Maroni~P, Robert~A 
1978b {\it Ann. Soc. Sci. Bruxelles} {\bf 92} 151--89
\item[] Erd\'elyi A 1942 {\it Duke Math. J.} {\bf 9} 48--58 
\item[] Erd\'elyi A 1944 {\it Q. J. Math.} {\bf 15} 62--9
\item[] Erd\'elyi, Magnus W, Oberhettinger F and Tricomi F G 1953
        \textit{Higher transcendental functions}
        (New York: McGraw-Hill)
\item[] Gautschi W 1967 {\it SIAM Rev.} {\bf 9} 24--82
\item[] de Gennes P G 1974 {\it J. Chem. Phys.} {\bf 60} 5030--42
\item[] Gerashchenko~S and Steinberg~V 2006 \textit{Phys. Rev. Lett.} 
        \textbf{96} 038304
\item[] Hatfield J W and Quake S R 1999 \textit{Phys. Rev. Lett.} 
        \textbf{82} 3548--51
\item[] Hern\'andez Cifre J G and Garc\'{\i}a de la Torre J 1998 
\textit{J. Non-Cryst. Solids} \textbf{235-237} 717--22
\item[] Hern\'andez Cifre J G and Garc\'{\i}a de la Torre J 1999 {\it J. Rheol.}
\textbf{43} 339--58
\item[] Hur~J~S, Shaqfeh~E~S, Babcock~H~P and Chu~S 2002 {\it Phys. Rev.} E
{\bf 66} 011915
\item[] Jaff\'e G 1933 {\it Z. Phys.} {\bf 87} 535--44
\item[] Larson R G, Perkins T T, Smith D E and Chu S \textit{Phys. Rev.} E
\textbf{55} 1794--7
\item[] Larson R G 2005 {\it J. Rheol.} {\bf 49} 1--70 
\item[] Leaver~E~W 1985 {\it Proc. R. Soc. London} A {\bf 402} 285--98
\item[] Manneville S, Cluzel Ph, Viovy J-L, Chatenay D and Caron F 1996
        \textit{Europhys. Lett.} \textbf{36} 413--8
\item[] Martins Afonso M and Vincenzi D 2005 {\it J. Fluid Mech.} {\bf 540}
        99--108
\item[] Morse P M and Feshbach H 1953 {\it Methods of Theoretical Physics} 
        (New York: McGraw--Hill)
\item[] Perkins T T, Quake S R, Smith D E and Chu S 1994 \textit{Science}
        \textbf{264} 822
\item[] Perkins T T, Smith D E and Chu S 1997 {\it Science} {\bf 276} 2016--21
\item[] Puliafito A and Turitsyn K 2005 {\it Physica}~D {\bf 211} 9--22
\item[] Quake S R, Babcock H and Chu S 1997 \textit{Nature} \textbf{388} 151--4
\item[] Rouse P E 1953 \textit{J. Chem. Phys.} \textbf{21} 1272--80 
\item[] Rzehak R and Zimmermann W 2002 \textit{Europhys. Lett.} \textbf{59}
        779--85
\item[] Schenzle A and Brand H 1979 {\it Phys. Rev.} A {\bf 20} 1628--47
\item[] Schroeder C M, Babcock H P, Shaqfeh E S G and Chu S 2003
{\it Science} {\bf 301} 1515--9
\item[] Schroeder C M, Shaqfeh E S G and Chu S 2004 {\it Macromolecules}
{\bf 37} 9242--56
\item[] Shaqfeh E S G 2005 \textit{J. Non-Newton. Fluid Mech.} \textbf{130}
1--28
\item[] Slavyanov S Yu 1995 {\it Confluent Heun equation}, in {\it Heun's Differential Equations} ed A. Ronveaux  (New York: Oxford University Press)
\item[] Slavyanov S Yu and Lay W 2000
{\it Special Functions, A Unified Theory Based on Singularities}
(New York: Oxford University Press)
\item[] Smith~D~E and Chu~S 1998 {\it Science} {\bf 281}, 1335--40
\item[] Smith D E, Babcock~H~P and Chu~S 1999 {\it Science} {\bf 283} 1724--7
\item[] Sreenivasan K R and White C M 2000 {\it J. Fluid Mech.} \textbf{409}
149--64
\item[] Stratonovich R L 1963 {\it Topics in the Theory of Random Noise} 
(New York: Gordon and Breach Science Publishers)
\item[] Svartholm N 1939 {\it Math. Ann.} {\bf 116}  413--21
\item[] Szeg\"o G 1939 {\it Orthogonal Polynomials} 
(Providence, RI: American Mathematical Society)
\item[] Wiest J L, Wedgewood L and Bird R B 1989 \textit{J. Chem. Phys.} \textbf{90} 587--94
\item[] Wimp~J 1984 {\it Computation with Recurrence Relations}
(Pitman Publishing Inc.)
\item[] Zimm~B~H 1956 {\it J. Chem. Phys.} \textbf{24} 269--78
\end{itemize}
\end{document}